\documentstyle[rmp,aps]{revtex}
\begin{document}
\title{Geometric aspects in Equilibrium Thermodynamics }
\author{L.Vel\'{a}zquez $^{1}$, F.Guzm\'{a}n $^{2}$}
\address{$^{1}$Departamento de Fisica, Universidad de Pinar del Rio, Cuba\\
e-mail: luisberis@geo.upr.edu.cu \\
$^{2}$Departamento de Fisica Nuclear, Instituto Superior de Ciencias y\\
Tecnolog\'{i}as Nucleares, Habana, Cuba.\\
e-mail: guzman@info.isctn.edu.cu }
\maketitle

\begin{abstract}
We discuss different aspects of the present status of the Statistical
Physics focusing the attention on the non-extensive systems, and in
particular, on the so called small systems. Multimicrocanonical Distribution
and some of its geometric aspects are presented. The same could be a very
atractive way to generalize the Thermodynamics. It is suggested that if the
Multimicrocanonical Distribution could be equivalent in the Thermodynamic
Limit with some generalized Canonical Distribution, then it is possible to
estimate the entropic index of the non-extensivethermodynamics of Tsallis
without any additional postulates.
\end{abstract}

\section{Introduction}

Traditionally, the Statistical Physics and the standard Thermodynamics have
been formulated to be applied to the study of extensive systems, in which
the consideration of the additivity and homogeneity properties, as well as
the realization of the Thermodynamical Limit are indispensable ingredients
for their performance. The study of these systems is based on the
consideration of the extensive Shannon-Boltzmann-Gibbs entropy $\left[ 1%
\right] $: 
\begin{equation}
S_{E}=-\sum_{k}p_{k}\ln p_{k}
\end{equation}
which satisfied the additivity and concavity conditions. This definition is
the base of the well known H theorem of Boltzmann, which is connected with
the Second Law of the Thermodynamic $\left[ 2\right] $, the law of the
growth of the entropy. This definition leads, during the thermodynamical
equilibrium, to the description of such systems by means of Boltzmann-Gibbs
Distributions$\left[ 1\right] $: 
\begin{equation}
\omega _{B-G}\left( X;I,N,a\right) =\frac{1}{Z\left( \beta ,N,a\right) }\exp %
\left[ -\beta \cdot I_{N}(X;a)\right]
\end{equation}

In fact, by means of this formalism, it is possible to obtain sound results
when it is applied to physical systems in which the above mentioned
conditions are valid i.e., to systems satisfying the homogeneity and
additivity conditions. However, these conditions can not be inferred from
general principles. Nowadays, \ these properties can be considered as
reasonable approximations for systems containing many particles interacting
by means of short range forces$\left[ 3-5\right] $, being these systems
homogeneous from the macroscopic point of view, so that they are not
suffering a phase transition of first order $\left[ 6,7\right] $. Out of
this context, the applicability of this theory arouse to be doubtful.

In the last years considerable efforts have been devoted to the study of
non-extensive systems. The available and increasingly experimental evidences
on anomalies presented in the dynamical and macroscopical behavior in plasma
and turbulent fluids $\left[ 8-11\right] $, astrophysical systems $\left[
12-17\right] $, nuclear $\left[ 18,19\right] $ and atomic clusters $\left[ 20%
\right] $, granular matter $\left[ 21\right] $, glasses$\left[ 22,23\right] $
and complex systems $\left[ 24\right] $, constitute a real motivation for
the generalization of the Thermodynamics.

\section{Recent Developments}

A step forward in these studies have been advanced in 1988 with the proposal
of Constantino Tsallis $[25]$ (see also $[26,27]$). \ In $[25]$ he
considered a generalization of the Boltzmann-Gibbs formalism to take into
account the non-extensive characteristics of a system introducing the
non-extensive entropy, {\it q-entropy}, $Sq$: 
\begin{equation}
S_{q}=\frac{1-\sum_{k}p_{k}^{q}}{q-1}
\end{equation}

In this case, $q$ , the entropic index, is a parameter that describes the
non extensive character of the system. It can be demonstrated that when $q$
approaches to the unit the q-entropy becomes in the traditional extensive
one. In the case of the systems in thermodynamic equilibrium, this
definition generalizes the Boltzmann-Gibbs distributions: 
\begin{equation}
\omega _{q}\left( X;I,N,a\right) =\frac{1}{Z_{q}\left( \beta ,N,a\right) }%
\left[ 1-\left( 1-q\right) \beta \cdot I_{N}(X;a)\right] ^{\frac{1}{1-q}}
\end{equation}
substituting this way the exponential laws for potential laws.

It is important to remark that with this kind of description one can obtain
good results when it is applied to a broad variety of non extensive and
complex systems. For example, in the case of non lineal systems, this
concept has leaded to generalize the Sinai-Komolgorov entropy, allowing the
study of chaotic regimens that could not be well described by other
formalisms $\left[ 28\right] $. However, one should mention that the
formalism can be classified as a parametrical one, since q can not be
estimated a priori for a given system and can not be derived from general
principles. In fact, it acts as an adjustment parameter.

In the last years there are important progresses in studying small systems,
understanding for such, those systems whose interaction range is long
comparable to or longer than the linear dimensions of the system. Under this
denomination can be grouped the molecular and atomic clusters and even the
astrophysical systems, confined under their own gravity, among others.

For such systems, the postulates of the traditional statistical physics do
not take place, because they are not homogeneous, their integrals of
movements are non additive, and they can not necessarily be considered as
systems of many particles. On the other hand, it is generally accepted, in
the framework of the standard theory, that the phase transitions only can
take place in systems which reach the thermodynamic limit, composed by a
huge number of particles. The study of the properties of the finite nuclear
matter has revealed experimental evidences of the occurrence of phase
transitions of first order during the nuclear multifragmentation $\left[ 29%
\right] $, forcing us to change our conceptions in this respect.

It is not difficult to notice of the great proximity that has the
no-extensive formulation of the thermodynamic given by Tsallis with their
antecedent, the Standard Thermodynamic:

\begin{itemize}
\item  Both are based on a probabilistic interpretation for the entropy,
characteristic that allows a great versatility for the application of this
important concept to other fields.

\item  A common denominator of both theories is the consideration of the
thermodynamic limit, the limit of many particles. In this way, the
description of the macroscopic state of \ equilibrium systems is carried out
by means of the parameters of the generalized canonical distributions and
therefore paying a full attention to the validity of the zero principle of
the Thermodynamics.
\end{itemize}

Here the reason that these theories can not be applicable to those systems
composed by few bodies, and in general, to the small systems. If we want to
arrive to a more general statistical formulation, it is necessary to abandon
definitively the postulates of the traditional theory. In this line, there
are important contributions of \ Prof. Gross at Hanh Meitner Institute of
Berlin who developed a formalism for analyzing the nuclear
multifragmentation \ and the statistical mechanics of small systems by means
of the Microcanonical Thermostatistics$\left[ 30,31\right] $.

To reach their objective, the same one returned at the pre-Gibbsianns times,
reconsidering the entropy concept given by Boltzmann, the celebrated epitaph
of his gravestone in Vienna: 
\begin{equation}
S_{B}\left( E,N,a\right) =\ln \left( W\left( E,N,a\right) \right)
\end{equation}
where: 
\begin{equation}
W\left( E,N,a\right) =\Omega \left( E,N,a\right) \delta \epsilon =\int
dX\delta \left[ E-H_{N}\left( X;a\right) \right] \delta \epsilon
\end{equation}
with the consideration of the ordinary Microcanonical Distribution for the
closed systems: 
\begin{equation}
\omega _{M}\left( X;E,N,a\right) =\frac{1}{\Omega \left( E,N,a\right) }%
\delta \left[ E-H_{N}\left( X;a\right) \right]
\end{equation}

From this base, the deduction of the Thermodynamic can only be developed
starting from the principles of the mechanics, without appealing to other
considerations. At this point is necessary to stress the hierarchy of the
microcanonical ensemble regarding the other statistical ensembles as pointed
out by Boltzmann$\left[ 32\right] $, Gibbs$\left[ 7\right] $, Einstein$\left[
33,34\right] $, Ehrenfests$\left[ 35\right] $ because the Canonical and
Grand Canonical ensembles can be derived from the Microcanonical one through
of certain particular conditions: {\it the extensity postulates}.

The meaning of the Boltzmann entropy is very direct, physically palpable,
valid for any system of particles, independently whether it is formed of
many particles or not: it is equal to the logarithm of the volume of the
hypersurface of constant energy in the N-body phase space of the system. By
this way, the Boltzmann entropy is a bridge, a direct connection between the
microscopic characteristics of the system with their macroscopic properties.

Nevertheless the achievements of the formalism, exist reasons to believe
that this is not still a completed one:

\begin{itemize}
\item  It is characteristic for this formulation the relevance of the
ergodic chauvinism of the traditional distributions. This is translated as
the preference of the energy on all the integral of movement corresponding
to the system.

\item  The consideration of the Boltzmann entropy is limited to the
equilibrium and therefore this important concept can not be extended to the
kinetics$\left[ 36\right] $.
\end{itemize}

Many of the systems traditionally described by the Statistical Physics are
considered located in certain region of the space due to the action of a
external confining field. It is the reason to consider and justify this
preference since among the integral of movements (the energy, the impulse
and total angular moment), the energy \ is only one conserved.

However, there is some systems in which the energy is not enough to describe
their macroscopic state and it is also necessary the consideration of the
total angular moment and other integral of movement in general . This
necessity becomes evident when we tried to make the macroscopic description
of the nucleus, as well as when we deal with the astrophysical systems$\left[
37,38\right] $.

These are the antecedents of the which starts the formulation of the
Multimicrocanonical Distribution$\left[ 39,40\right] $. As their precedent,
this formulation pretend to be essentially geometric, characteristic that
not only could facilitate handling this distribution, but it could have also
deeper consequences in the understanding of the Statistical Physics of the
systems in equilibrium thermodynamics .

\section{Geometrical Aspects of the Multimicrocanonical Distribution}

Let $\left( I,N,a\right) $ represent the macroscopic state of a system,\
where $I\equiv \left\{ I^{1},I^{2}\ldots I^{n}\right\} $\ are the set of
integrals of movement of the distribution, $N$\ \ the particle number and $a$%
\ describe an external parameter. Following the general ideas of the
Microcanonical Thermostatistic, the probability phase density is given
by:\bigskip 
\begin{equation}
\omega _{M}\left( X;I,N,a\right) =\frac{1}{\Omega \left( I,N,a\right) }%
\delta \left( I-I_{N}\left( X;a\right) \right)
\end{equation}
generalizing the ordinary microcanonical distribution, where $\Omega $\ is
the state density. The choice of equal-probability hypothesis allows to take
a totally geometrical version for the statistic mechanics without the
introduction of artificial elements in the formulation.

Similarly, we assumed the Boltzmann's definition of entropy:

\begin{equation}
S_{B}\left( I,N,a\right) =\ln \left( \Omega \left( I,N,a\right) \delta
I\right)
\end{equation}
where $\delta I$ is a suitable elemental volume. The Boltzmann definition of
entropy does not require to satisfy the concavity and extensity conditions
that ordinary exhibits the information entropy. How we can see, this
definition is essentially geometrical and transparent in its sense.

As we showed in Ref. $\left[ 40\right] $, in this formulation appears the
concepts of scalar product and divergency of vectors belonging \ to a
geometrical approach. It motived to develop a geometrical formalism to work
with this distribution. The geometrical aspects of the theory are summed up
by the fallowing facts:

\begin{itemize}
\item  Any function of the movement integrals is itself a movement integral.
If we have a complete independent set of functions of the movement integral
of the distribution, then this set of functions is equivalent to the other
set of movement integrals, representing the same macroscopic state of the
system. That is why it is more exactly to speak about an abstract Space of
movement integrals for the multimicrocanonical distribution, $\Im _{N}$.
Therefore, every physical quantity or behavior has to be equally reproduced
by any representation of the space of macroscopic state, $\left( \Im
_{N};a\right) $\ .

\item  It is easy to show that the multimicrocanonical distribution is local
reparametrization invariant. Let $\Re _{I}$ and $\Re _{\phi }$ $\left( \
I\equiv \left\{ I^{1},I^{2}\ldots I^{n}\right\} \ \text{, }\ \phi \equiv
\left\{ \phi ^{1}\left( I\right) ,\phi ^{2}\left( I\right) \ldots \phi
^{n}\left( I\right) \right\} \right) $\ \ two representations of $\Im _{N}$.
By the property of the $\delta $-function we have:
\end{itemize}

\begin{equation}
\delta \left[ \phi \left( I\right) -\phi \left( I_{N}\left( X;a\right)
\right) \right] =\left| \frac{\partial \phi }{\partial I}\right| ^{-1}\delta %
\left[ I-I_{N}\left( X;a\right) \right]
\end{equation}

hence:

\begin{equation}
\widetilde{\Omega }\left( \phi ,N,a\right) =\left| \frac{\partial \phi }{%
\partial I}\right| ^{-1}\Omega \left( I,N,a\right)
\end{equation}

and therefore:

\begin{equation}
\omega _{M}\left( X;\phi ,N,a\right) =\omega _{M}\left( X;I,N,a\right)
\equiv \omega _{M}\left( X;\Im _{N},a\right)
\end{equation}

\begin{itemize}
\item  \ The state density allows to define the invariant measure of the
space : 
\begin{equation}
d\mu =\Omega \left( I,N,a\right) d^{n}I
\end{equation}
although this is the most important measure for the space, this is not the
only one, because there are others like them derived from Poincare
invariants, when we project the N-body phase-space to the space\ $\Im _{N}$.

\item  The movement integrals are defined by the commutativity relation with
the Hamiltonian of the system. In the case of closed systems, the
Hamiltonian is the energy of the system, and this is a conserved quantity.
In the multimicrocanonical distribution it represents one of the coordinate
of the point belonging to $\Im _{N}$, in an specific representation. When we
change the coordinate system, the energy lose its identity. Since we can not
fix the commutativity of the energy with the others integrals, it would be
more correct that all movement integrals commute between them in order to
preserve these conditions. Thus, we make consistently the local
reparametrization invariance with the commutativity relations. As we see,
still from the classical point of view, the reparametrization invariance
suggests that the set of movement integrals have to be\ simultaneously
measured. In the quantum case, this is an indispensable request for the
correct definition of the statistical distribution. This property is landed
to the classical distribution by the correspondent principle between the
Quantum and the Classic Physic.
\end{itemize}

This reparametrization freedom is very attractive: we can choose the
representation of the macroscopic state space more adequate to describe the
statistical system. There are many examples in which an adequate choice of
coordinate system helps us to simplify the resolution of a problem.

Let us move now to the Boltzmann definition of entropy. Obviously\ it has to
be an scalar function. Its value can not depends on the coordinate system
used to describe the macroscopic state. It is our interest that this
function characterizes the macroscopic state of the system, it have to allow
us to get information about the ordering of it. In this case, $\delta I$ can
not be arbitrary. It has to be a characteristic of the representation of the
space $\Im _{N}.$

These facts aim toward the creation of a statistical theory conceived as a
local geometric theory. To achieve this it should be endowed to the space $%
\Im _{N}$ of a geometric structure: we have to introduce a theory of tensors
and vectors and incorporate the covariant differentiation , local bases,
etc. We are interested in describing those properties of system possessing
topological and geometrical invariance, since they represent characteristics
illustrating the behavior of the system.

\section{General perspectives.}

By taking into account the previous discussions, the problem is reduced to
endow the space of the movement integrals of the system, $\Im _{N}$ , of a
geometric structure. When we speak on a geometric structure, we are
referring to consider it as a topological and metric space. This aspect is
not new in the statistical physics. An antecedent of this approach is the
geometric interpretation make by Weinhold\ $\left[ 41\right] $\ to the
Standard Thermodynamic for extensive systems. However, this approach is not
valid for a more general case.

A paradigmatic point in this direction is the supposition of the existence
of a internal product among vectors in this space, it means, the
consideration of a metric tensor $g_{\mu \nu }$. It constitutes the basis
for the definition of invariant quantities and of the covariant
differentiation$\left[ 42\right] $.

Having in mind the fundamental character that this concept has for the
theory, it is not difficult to admit that a possible and reasonable
generalization of the Boltzmann entropy should be: 
\begin{equation}
S_{B}=\ln \left[ \frac{\Omega \left( I,N,a\right) }{\sqrt{\left| g\right| }}%
\right]
\end{equation}
where $g=\det \left( g_{\mu \nu }\right) $. In this way, the Boltzmann
entropy results an scalar function, allowing the characterization of the
macroscopic state of the system.

Are very well-known the difficulties presented during the initial
development of the statistical mechanics, because it was a theory based in
the microscopics laws of the classic physics, a theory on the continuous.
These difficulties were overcome only with the arriving of the quantum
Physics. It is very well-known too how Boltzmann appeals to the idea of the
quantification in order to deal with a well defined definition of entropy ,
consideration that years later was consider by Planck in the interpretation
of the very celebrated formula for the of emission spectrum of the black body%
$\left[ 43\right] $. This is a very interesting connection between the
quantum theory with the statistical mechanics.

The consideration of a geometric structure for the space $\Im _{N}$ is an
alternative for the correct definition of the statistics. The metric tensor
should be derived from magnitudes that have their origin in the principles
of the mechanics. The same one should be derived starting from certain
structural equations whose form we still ignores. However, for the
completeness of the theory, these equations should be intimately bond to the
quantum properties of the systems.

An essentially important fact for this theory is that the same one presents
two general symmetries: it is invariant under the {\it local
reparametrizations or diffeomorfism} of $\Im _{N}$, $Diff(\Im _{N})$ , and
under the transformations of the {\it unitary lineal group}, $SL(n,R)$.

It is necessary to recall that the first one constitutes the maximum
symmetry that a theory of geometric character can possess, associating it to
the multimicrocanonical distribution. On the other hand, the transformations
of the unitary lineal group $SL(n,R)$ constitutes the group of more general
symmetry that possess the distributions of Boltzmann-Gibbs and of the
derived potential distributions of the theory of Tsallis. In this case,
these transformations should respect the properties of scaling of the
movements integral for these distributions. To be more specific, in the
peculiar case of the distributions of Boltzmann-Gibbs these transformations
respect the additivity of the integrals of movement. This symmetry group
acts on the space vectorial tangent to each point of the space $\Im _{N}$,
i.e. , an euclidean space.

In the particular case of the extensive systems, if we considered the
thermodynamic limit in the representation of the space $\Im _{N}$ through
the additives integral of movement, the multimicrocanonical distribution
degenerates in the Boltzmann-Gibbs distributions. This fact, together with
the consideration of the thermodynamic limit lead to a rupture of the
symmetry from $Diff(\Im _{N})$ to $SL(n,R)$, from a curved to euclidean
space.

If we extrapolate this interpretation to the general case, the
Multimicrocanonical Distribution would become what we denominate a Parent
Theory for all the Statistical Distributions for the systems in macroscopic
equilibrium when the thermodynamic limit is reached. The Boltzmann entropy
would allow a probabilistic interpretation to the style of Shannon
-Boltzmann-Gibbs entropy and the non extensive entropy of \ Tsallis. In the
last case, \ it is really interesting point because open the door to a
possibility to obtain the entropic index q without the necessity of
appealing to additional postulates.

\section{Conclusions}

The geometrical interpretation of the equilibrium thermodynamics is a very
interesting way to generalize it. The Boltzmann definition of entropy is the
most transparent connection between the Microscopical and Statistical
Mechanics, and it is probable the only way to extended the thermodynamics to
the few-body systems. If the Multimicrocanonical Distribution could be
equivalent in the Thermodynamic Limit with some generalized Canonical
Distribution, then it is possible to estimate the entropic index of the
non-extensive thermodynamics of Tsallis without any additional postulates. \
\ \ \ \ \ \

\end{document}